\documentclass[%
 reprint, twocolumn,
 amsmath,amssymb,
 aps,
]{revtex4-2}
\usepackage{}

\usepackage{graphicx}
\usepackage{dcolumn}
\usepackage{bm}
\usepackage{epstopdf}
\usepackage{float}
\usepackage{xcolor, colortbl}
\usepackage{hyperref}

\newcommand{\bsub}{\begin{subequations}}
\newcommand{\esub}{\end{subequations}}
\newcommand{\lrb}[1]{\left(#1\right)}
\newcommand{\lrs}[1]{\left[#1\right]}

\begin{document}

\preprint{APS/123-QED}

\title{Coupling shape and pairing vibrations in a collective Hamiltonian based on nuclear energy density functionals (II):   low-energy excitation spectra of triaxial nuclei}

\author{J. Xiang}
\affiliation{College of Physics and Electronic Engineering, Chongqing Normal University, Chongqing 401331, China}
\affiliation{School of Physics and Electronic, Qiannan Normal University for Nationalities, Duyun, 558000, China}
\author{Z. P. Li}
\email{zpliphy@swu.edu.cn}
\affiliation{School of Physical Science and Technology, Southwest University, Chongqing 400715, China}

\author{T. Nik\v si\' c}
\affiliation{Physics Department, Faculty of Science, University of Zagreb, Croatia}
\author{D. Vretenar}
\affiliation{Physics Department, Faculty of Science, University of Zagreb, Croatia}

\author{W. H. Long}
\affiliation{School of Nuclear Science and Technology, Lanzhou University, Lanzhou 730000, China}

\author{X. Y. Wu}
\affiliation{College of Physics and Communication Electronics, Jiangxi Normal University, Nanchang 330022, China}

\date{\today}

\begin{abstract}
The triaxial quadrupole collective Hamiltonian, based on relativistic energy density functionals, is extended to include a pairing collective coordinate. In addition to triaxial shape vibrations and rotations, the model describes pairing vibrations and the coupling between triaxial shape and pairing degrees of freedom. The parameters of the collective Hamiltonian are determined by a covariant energy density functional, with constraints on the intrinsic triaxial shape and pairing deformations. The effect of coupling between triaxial shape and pairing degrees of freedom is analyzed in a study of low-lying spectra and transition rates of $^{128}$Xe. When compared to results obtained with the standard triaxial quadrupole collective Hamiltonian, the inclusion of dynamical pairing compresses the low-lying spectra and improves interband transitions, in better agreement with data. The effect of zero-point energy (ZPE) correction on low-lying excited spectra is also discussed.
\end{abstract}

\maketitle


\section{\label{sec:I}Introduction}

The occurrence of pairing vibrations in nuclei was suggested by Bohr and Mottelson in the 1960s \cite{Bohr1964}, and this mode influences many physical quantities, such as low-lying excitation spectra \cite{KyotokuPRC1987,VaqueroPLB2011,VaqueroPRC2013}, two-nucleon transfer reactions \cite{SiegalNPA1972}, nuclear matrix elements for neutrinoless double-beta decay \cite{NurialopezPRL2013}, spontaneous fission half-lifes \cite{StaszczakNPA1989,SadhukhanPRC2014,ZhaoPRC2016,GiulianiPRC2014,RodriguezPRC2018}, and induced fission yields \cite{JZhao2021PRC}. In particular, for low-lying spectra, a number of pairing vibrational states have been observed in heavier nuclei, e.g. the proton pairing vibrational states in $^{208}$Pb \cite{IGO1971AnP,Anderson1977PRL}, $^{206}$Pb \cite{Anderson1977PRL}, $^{124}$Xe, and $^{126}$Xe \cite{AlfordNPA1979,RadichPRC2015}.

A variety of theoretical methods have been used to describe pairing vibrations: the pairing Hamiltonian \cite{BesNP1966,BrogliaNPA1967,BrogliaNPA1968}, the collective Hamiltonian \cite{BESNPA1970,DusselNPA1971,DusselNPA1972,zajacAPPSB1999,ProchniakNPA1999,ZajacNPA1999,zajacAPPB2000,Prochniak2002APPB,SrebrnyNPA2006,pomorskiPhS2000,ZajacAPPB2001}, time-dependent Hartree-Fock-Bogoliubov (TDHFB) theory \cite{AvezPRC2008,RipkaNPA1969}, the shell model \cite{HeuslerPRC2015}, the quasiparticle random phase approximation \cite{HirotakaPRC2011,KhanPRC2009,SharmaPRC1994,Dussel1988PRC,BesNP1966,BrogliaNPA1968b,GoswamiPRC1970}, the pair-addition and pair-removal phonon model \cite{JolosPRC2015}, and the generator coordinate method (GCM) \cite{RipkaNPA1969,SiegalNPA1972,FaesslerNPA1973,GozdzNPA1985,PomorskiAPPB1987,SiejaEPJA2004,ProchniakIJMPE2007,PomorskiIJMPE2007}. In general, however, these methods have not explicitly considered the coupling between shape and pairing vibrations.

In a recent study \cite{XiangPRC2020}, we have extended the quadruple collective Hamiltonian (QCH) to include a pairing collective coordinate. The quadrupole-pairing collective Hamiltonian (QPCH) is based on the framework of covariant density functional theory (CDFT). In addition to the quadrupole shape vibrations and rotations, the model describes pairing vibrations and explicitly couples shape and pairing degrees of freedom. It has been shown that the inclusion of dynamical pairing increases the moment of inertia and collective mass, lowers the energies of excited $0^+$ states and bands built on them, reduces the $E0$ transition strengths and, generally, produces low-lying spectra in much better agreement with experimental results.

The breaking of axial symmetry in the nuclear intrinsic state influences both structural and dynamical properties. In general, it leads to an increase of binding energies \cite{MollerPRL2006}, a lowering of fission barriers in heavy nuclei \cite{LiPRC2010f,LuPRC2012}, brings low-lying excitation spectra of shape-coexisting nuclei in better agreement with data \cite{FuPRC2013,LiJPG2016,YangPRC2021}, leads to the onset of wobbling motion \cite{BohrBook1975} and nuclear chirality \cite{FrauendorfNPA1997,MengPRC2006}. As emphasized in Ref. \cite{XiangPRC2020}, the effect of pairing vibrations will be particularly important for $\gamma$-soft nuclei characterized by shape coexistence \cite{HeydeRMP2011} and, therefore, it is important to develop a model that allows for the coupling between pairing and triaxial ($\beta,\gamma$) shape degrees of freedom. In Refs. \cite{NomuraPRC2021a,NomuraPRC2021b}, we have attempted to describe the coupling of pairing and triaxial shape vibrations in collective states of $\gamma$-soft nuclei in the framework of the interacting boson model (IBM) mapped from the CDFT deformation energy manifold. Illustrative calculations for Xe, Os, and Pt isotopes show that, by simultaneously considering both shape and pairing collective degrees of freedom, the CDFT-based IBM successfully reproduces data on collective structures based on low-energy $0^+$ states, as well as $\gamma$-vibrational bands.

In the present work we develop a collective Hamiltonian that includes the degrees of freedom of pairing vibration, triaxial shape vibrations, and rotations. The inertia parameters and the collective potential are microscopically determined by the nuclear energy density functional. The theoretical framework is outlined in Sec. \ref{Sec:II}. In Sec. \ref{Sec:III} we calculate  the low-lying spectra and transition rates of $^{128}$Xe. Section \ref{Sec:IIII} presents a brief summary of this study.


\section{\label{Sec:II}Theoretical Framework}

Nuclear excitations characterized by triaxial quadrupole shape vibrational and rotational collective motion, and  coupled with pairing vibrations, can be described by constructing a collective Hamiltonian defined by the quadrupole shape deformation parameters $\beta$ and $\gamma$, the Euler angles $\Omega$, and the pairing deformation $\alpha$ as collective coordinates (denoted as TPCH). The collective Hamiltonian takes the general form
\begin{align}
   \label{eq:TPCH}
{\hat H}_{\rm coll} &= -\frac{\hbar^2}{2\sqrt{\omega{r}}}\sum\limits_{i,j}\frac{\partial}{\partial q_i}\sqrt{\omega{r}}(B^{-1})_{ij}\frac{\partial}{\partial q_j}\nonumber\\
&+\frac{1}{2}\sum\limits_{k=1}^3\frac{\hat{J}_k^2}{{\cal I}_k}
+{V}_{\rm coll}(q),
\end{align}
where the collective mass tensor is defined by
\begin{align}
B=\left(
\begin{array}{ccc}
B_{\beta\beta} & \beta B_{\beta\gamma}     & B_{\beta\alpha} \\
\beta B_{\beta\gamma} & \beta^2B_{\gamma\gamma} &\beta{B}_{\gamma\alpha}\\
B_{\beta\alpha} & \beta B_{\gamma\alpha}   & B_{\alpha\alpha}\\
\end{array}
\right),
\end{align}
and $\omega={\rm det} B$. $r={\cal I}_1{\cal I}_2{\cal I}_3$, and the moments of inertia read
\begin{align}
{\cal I}_k=4B_k\beta^2\sin^2\lrb{\gamma-2k\pi/3}.
\end{align}


The entire dynamics of the collective Hamiltonian Eq.~(\ref{eq:TPCH}) is governed by the ten functions of the intrinsic quadrupole deformations $\beta$ and $\gamma$, and the pairing deformation $\alpha$: the collective potential, the six mass parameters $B_{\beta\beta}$, $B_{\gamma\gamma}$, $B_{\alpha\alpha}$, $B_{\beta\gamma}$, $B_{\beta\alpha}$, $B_{\gamma\alpha}$, and the three moments of inertia $\mathcal{I}_k(k=1,2,3)$. These functions are determined microscopically by constrained CDFT calculations. In the present study the energy density functional PC-PK1 \cite{ZhaoPRC2010} determines the effective interaction in the particle-hole channel, and the Bardeen-Cooper-Schrieffer (BCS) approximation with a separable pairing force is employed in the particle-particle channel \cite{TianPLB2009,NIksicPRC2010}. The framework of CDFT plus BCS with a separable pairing force is described in detail in Ref. \cite{Xiang2012NPA}.

The map of the collective energy surface as a function of $\beta$, $\gamma$, and $\alpha$ is obtained by imposing constraints on the mass quadrupole moments $q_{20}$, $q_{22}$, and pairing deformation $\alpha$, respectively \cite{RingBook,SiejaEPJA2004}.
\begin{align}
&\langle \hat H\rangle+\sum\limits_{\mu=0,2}C_{2\mu}\left(\langle \hat{Q}_{2\mu}\rangle - q_{2\mu}\right)^2\nonumber\\
&-\lambda\langle\hat{N}-N\rangle
-\xi_\alpha\langle\hat{P}-\alpha\rangle,
\label{constr}
\end{align}
where $\langle \hat H\rangle$ is the total energy, and  $\langle \hat{Q}_{2\mu}\rangle$ denotes the expectation value of the mass quadrupole operator:
\begin{align}
\hat{Q}_{20}=2z^2-x^2-y^2\ \  \text{and}\ \  \hat{Q}_{22}=x^2-y^2.
\end{align}
$q_{2\mu}$ is the constrained value of the quadrupole moment, and $C_{2\mu}$ the corresponding stiffness constant \cite{RingBook}. 
$\hat{N}$ is the particle number operator, while $\hat{P}$ is the pairing operator
\begin{align}
\hat{P} = \frac{1}{2}\sum\limits_{k>0}(c_kc_{\bar{k}}+c^\dagger_{\bar{k}}c^\dagger_{k}).
\end{align}
 $\lambda$ and $\xi_\alpha$ are Lagrange multipliers. $N$ and $\alpha$ are the constrained values of the particle number and pairing deformation, respectively.

The single-nucleon wave functions, energies and occupation probabilities, generated from constrained CDFT calculations, provide the microscopic input for the parameters of the collective Hamiltonian. The moments of inertia are calculated according to the Inglis-Belyaev formula \cite{InglisPR1956,BeliaevNP1961}
\begin{align}
\label{eq:MOI}
\mathcal{I}_k &= \sum_{i,j}{\frac{\left(u_iv_j-v_iu_j \right)^2}{E_i+E_j}
  | \langle i |\hat{J}_k | j  \rangle |^2},
\end{align}
where $k$ denotes the axis of rotation, and the summation runs over the proton and neutron quasiparticle states. The mass parameters are calculated in the cranking approximation \cite{GirodNPA1979}:
\begin{align}
B_{\mu\nu}=\hbar^2\lrs{\mathcal{M}^{-1}_{(1)}\mathcal{M}_{(3)}\mathcal{M}^{-1}_{(1)}}_{\mu\nu}
\end{align}
in which $\mu,\nu$ indicate $\alpha$, $q_{20}$ and $q_{22}$, with
\begin{align}
\mathcal{M}_{(n)\mu\nu}&=\sum\limits_{ij}\frac{\mathcal{O}^\mu_{ij}\mathcal{O}^\nu_{ji}}{\lrb{E_i+E_j}^n}\\
\mathcal{O}^{q_{20}}_{ij}&=\left\langle i\right|\hat{Q}_{20}\left| j\right\rangle
\left(u_i v_j+ v_i u_j \right)\\
\mathcal{O}^{q_{22}}_{ij}&=\left\langle i\right|\hat{Q}_{22}\left| j\right\rangle
\left(u_i v_j+ v_i u_j \right)\\
\mathcal{O}^\alpha_{ij}&=-\frac{1}{2}\lrb{u^2_{i}-v^2_{i}}\delta_{ij}
\end{align}

The collective potential $V_{\rm coll}$ in Eq.~(\ref{eq:TPCH}) is obtained by subtracting the vibrational and rotational zero-point energy (ZPE) corrections from the total mean-field energy:
\begin{align}
{V}_{\rm coll} =  \langle \hat H\rangle - \Delta V_{\rm vib} - \Delta V_{\rm rot}.
\end{align}
The vibrational ZPE corrections are calculated in the cranking approximation \cite{GirodNPA1979}
\begin{align}
\Delta{V}_{\rm vib}=\frac{1}{4}\text{Tr}\lrb{\mathcal{M}^{-1}_{(3)}\mathcal{M}_{2}}_{\mu\nu}.
\end{align}
The rotational ZPE is a sum of three terms:
\begin{align}
\Delta V_{\rm rot}&=\Delta{V}_{-2-2}\lrb{q_{20},q_{22},\alpha}+\Delta{V}_{-1-1}\lrb{q_{20},q_{22},\alpha}\nonumber\\
&+\Delta{V}_{11}\lrb{q_{20},q_{22},\alpha}
\end{align}
with
\begin{align}\label{eq:Deltamn}
\Delta{V}_{\mu\nu}\lrb{q_{20},q_{22},\alpha}=\frac{1}{4}
\frac{\mathcal{M}_{(2),\mu\nu}\lrb{q_{20},q_{22},\alpha}}{\mathcal{M}_{(3)\mu\nu}\lrb{q_{20},q_{22},\alpha}}
\end{align}
The individual terms are calculated from Eq.~(\ref{eq:Deltamn}),
with the intrinsic components of the quadrupole operator defined by
\begin{align}
\hat{Q}_{21}=-2iyz\ \ \hat{Q}_{2-1}=-2xz,\ \ \hat{Q}_{2-2}=2ixy.
\end{align}
The derivation of collective masses and the ZPE is explained in the Appendix.

The diagonalization of the collective Hamiltonian Eq.~(\ref{eq:TPCH}) yields the energy spectrum $E^I_i$ and the corresponding eigenfunctions
\begin{align}\label{eq-Wav}
\Psi^{IM}_n\lrb{\beta,\gamma,\alpha,\Omega}=\sum\limits_{K\in\Delta{I}}
\psi^{I}_{nK}\lrb{\beta,\gamma,\alpha}\Phi^I_{MK}\lrb{\Omega}.
\end{align}
Using the collective wave functions~(\ref{eq-Wav}), various observables
can be calculated and compared with experimental results. For
instance, the reduced electric quadrupole transition probability $B(E2)$ reads
\begin{align}
B\lrb{E2;nI\rightarrow{n^\prime}I^\prime}=\frac{1}{2I}\left|\langle{n^\prime}I^\prime\mid\mid\hat{\mathcal{M}}\lrb{E2}
\mid\mid{n}I\rangle\right|^2,
\end{align}
where $\hat{\mathcal{M}}$ is the quadrupole operator in the labrotatory frame. The electric monopole transition probability can be calculated from
\begin{align}
\rho^2\lrb{E0;nI\rightarrow{n^\prime I}}
=\left|\frac{\langle nI\mid\sum_ke_kr^2_k\mid n^\prime I \rangle}{eR^2_0}\right|^2,
\end{align}
with $R_0\simeq1.2A^{1/3}$ fm.
For a given collective state, the probability density distribution in the
$(\beta,\ \gamma,\ \alpha)$ plane is defined as
\begin{align}
\rho_{nI}\lrb{\beta,\gamma,\alpha}=\sum\limits_{K\in\Delta{I}}\left|\psi^I_{nK}\lrb{\beta,\gamma,\alpha}\right|^2
\beta^3\left|\sin{3\gamma}\right|
\end{align}
with the normalization
\begin{align}
  \label{eq:rho-nI}
  \int^\infty_0{\rm d}\alpha\int^\infty_0\beta{\rm d}\beta\int^{2\pi}_0\rho_{nI}\lrb{\beta,\gamma,\alpha}{\rm d}\gamma=1.
\end{align}


\section{\label{Sec:III}Results and discussions}

To illustrate and test the model that couples triaxial shape ($\beta,\gamma$) and pairing vibrations ($\alpha$), we calculate the constrained potential energy surfaces, the resulting collective excitation spectra and transition rates for $^{128}$Xe, a typical $\gamma$-soft nucleus. Following our previous study in Ref.~\cite{XiangPRC2020}, in the present CDFT calculation the strength parameter of the separable pairing force is enhanced by $6\%$ compared to the original value determined in Refs.~\cite{TianPLB2009,NIksicPRC2010}, namely here $G=-771.68\text{MeV}~\text{fm}^3$ is used.

\subsection{Effect of dynamical pairing and triaxial deformation on excitation spectra}

\begin{figure}[ht]
  \centering{\includegraphics[width=0.46\textwidth]{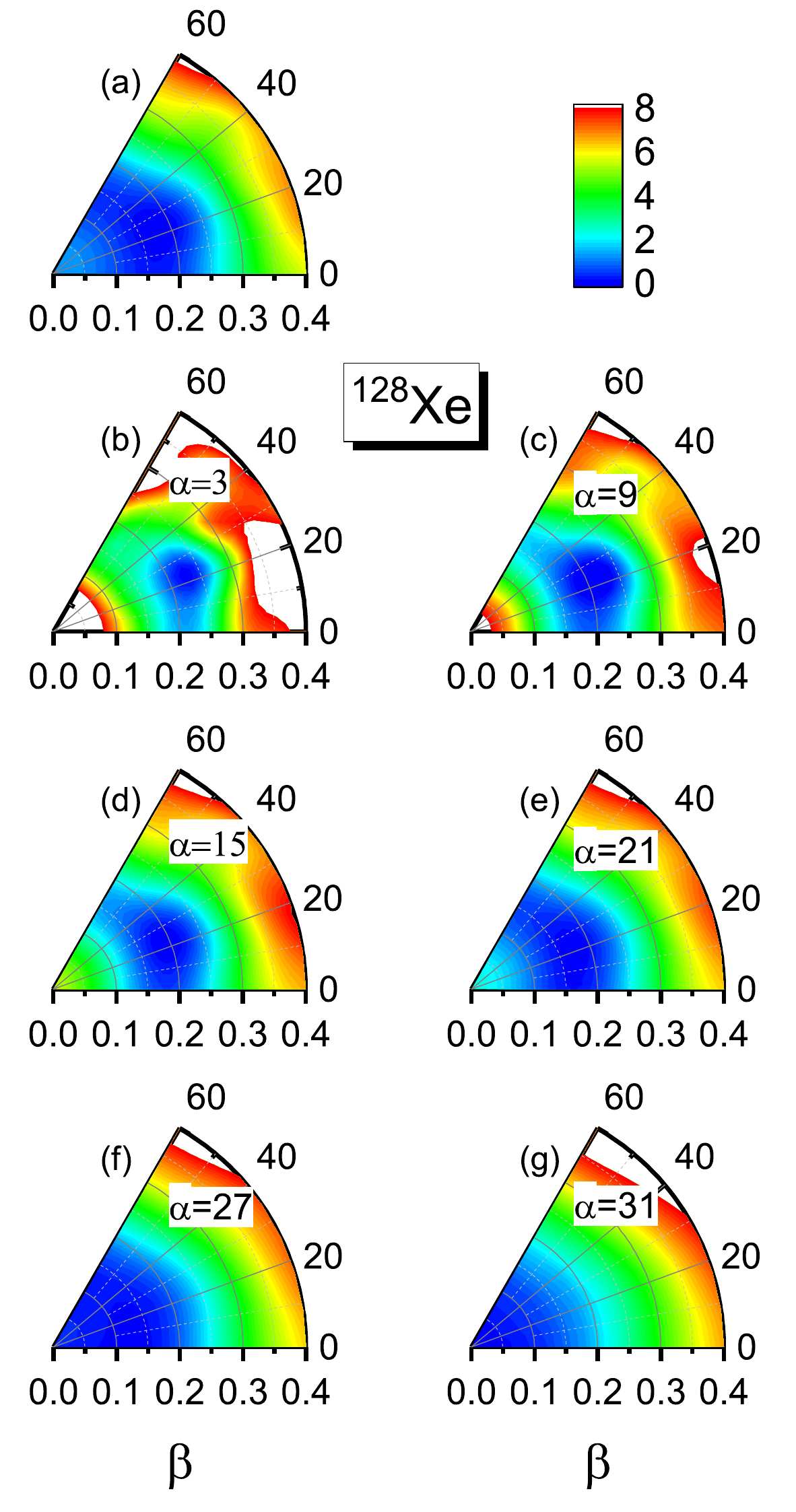}}
  \caption{\label{PES-beta-gamma}(Color online) The collective potential energy surfaces (PESs) in the ($\beta, \gamma$) plane calculated without and with constraints on $\alpha$ are shown in panel (a), and panels (b-g) $\alpha=3-31$, respectively. All energies (in MeV) are normalized with respect to the energy of the absolute minimum. The contours join points on the surface with the same values.}
  \end{figure}

  \begin{figure}[ht]
  \centering{\includegraphics[width=0.47\textwidth]{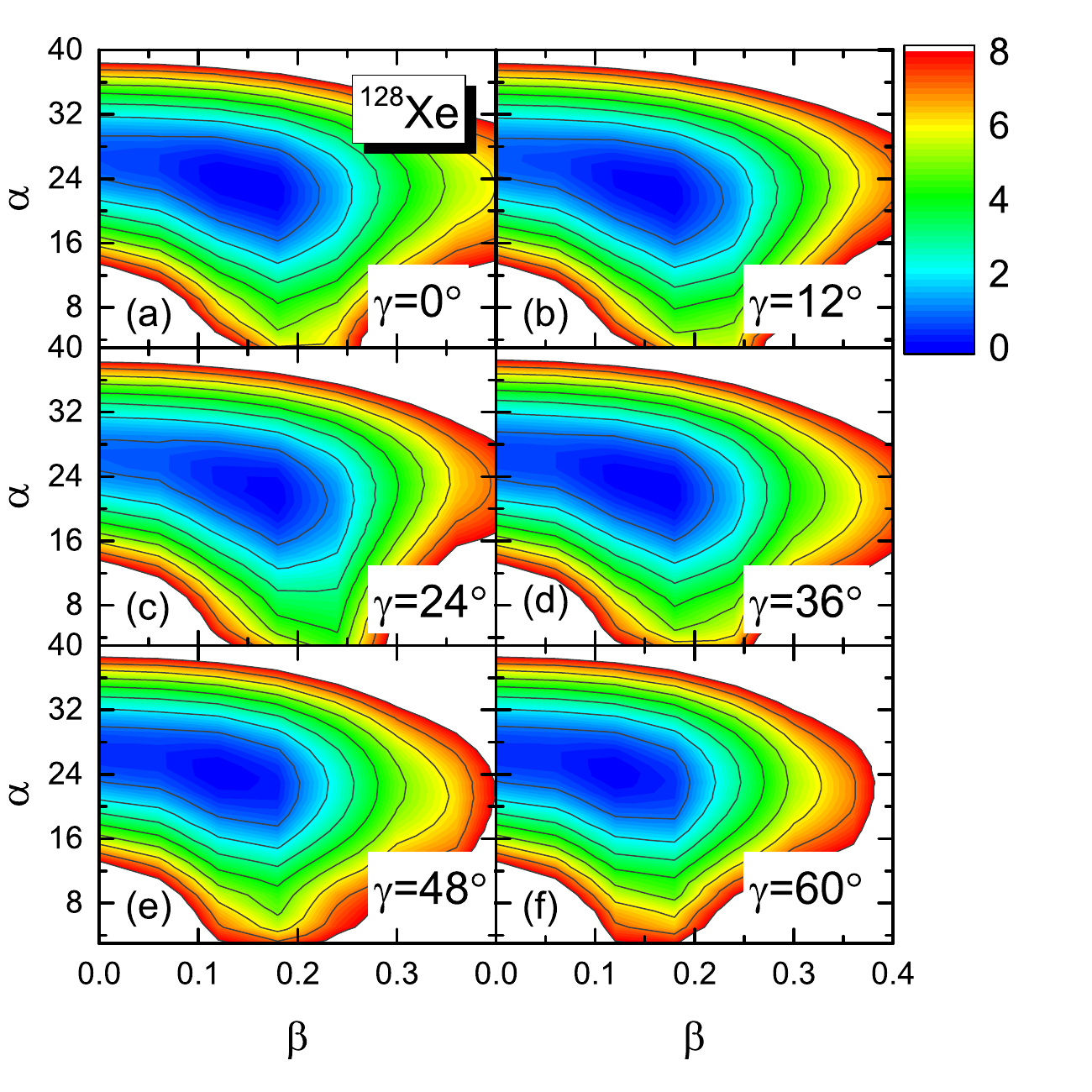}}
  \caption{\label{PES-beta-alpha}(Color online) Same as in the caption to Fig. \ref{PES-beta-gamma}, but for the PESs in the ($\beta, \alpha$) plane with constant $\gamma$ values.}
  \end{figure}

  \begin{figure}[ht]
  \centering{\includegraphics[width=0.47\textwidth]{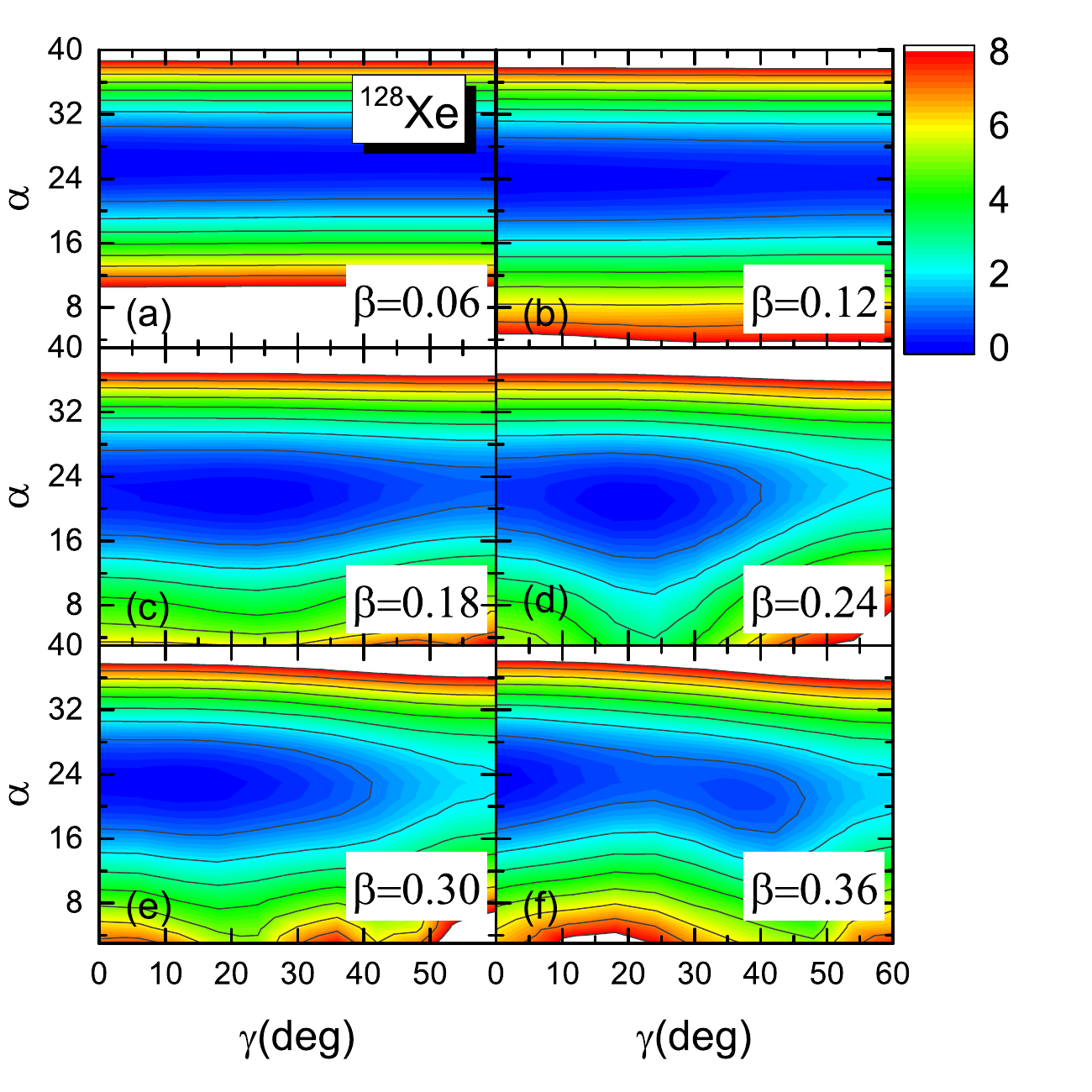}}
  \caption{\label{PES-gamma-alpha}(Color online) Same as in the caption to Fig. \ref{PES-beta-gamma}, but for the PESs in the ($\gamma, \alpha$) plane with constant $\beta$ values.}
  \end{figure}

The three-dimensional collective potential of $^{128}$Xe, obtained in the self-consistent CDFT calculation with constraints on $(\beta, \gamma, \alpha)$, is projected onto the corresponding two-dimensional planes in Figs. \ref{PES-beta-gamma}-\ref{PES-gamma-alpha}. Figure~\ref{PES-beta-gamma} displays the PESs in the ($\beta, \gamma$) plane calculated without and with constraints on $\alpha$, in panels (a) and (b-g), respectively. From panel (a), one notices that $^{128}$Xe is $\gamma$-soft with a shallow global minimum at $(\beta,\ \gamma,\ \alpha)\approx(0.18,\ 18^\circ,\ 21)$. As the paring changes from weak ($\alpha=3$) to strong ($\alpha=31$), the PESs display a very interesting evolution, that is, from triaxial to $\gamma$-soft, then to soft in both $\beta$ and $\gamma$, and finally to spherical. The PESs in the ($\beta, \alpha$) plane exhibit a similar pattern, soft for $\beta<0.2$ and $\alpha\sim 24$,  while the shape varies from prolate ($\gamma=0^{\rm o}$) to oblate ($\gamma=60^{\rm o}$). Remarkably, in Fig. \ref{PES-gamma-alpha} it is shown that the PESs remains $\gamma$-soft for a large interval of $\beta$ and $\alpha$ values: $\beta\leq 0.36$ and $16 \leq\alpha\leq 32$.

\begin{figure}[ht]
  \centering{\includegraphics[width=0.48\textwidth]{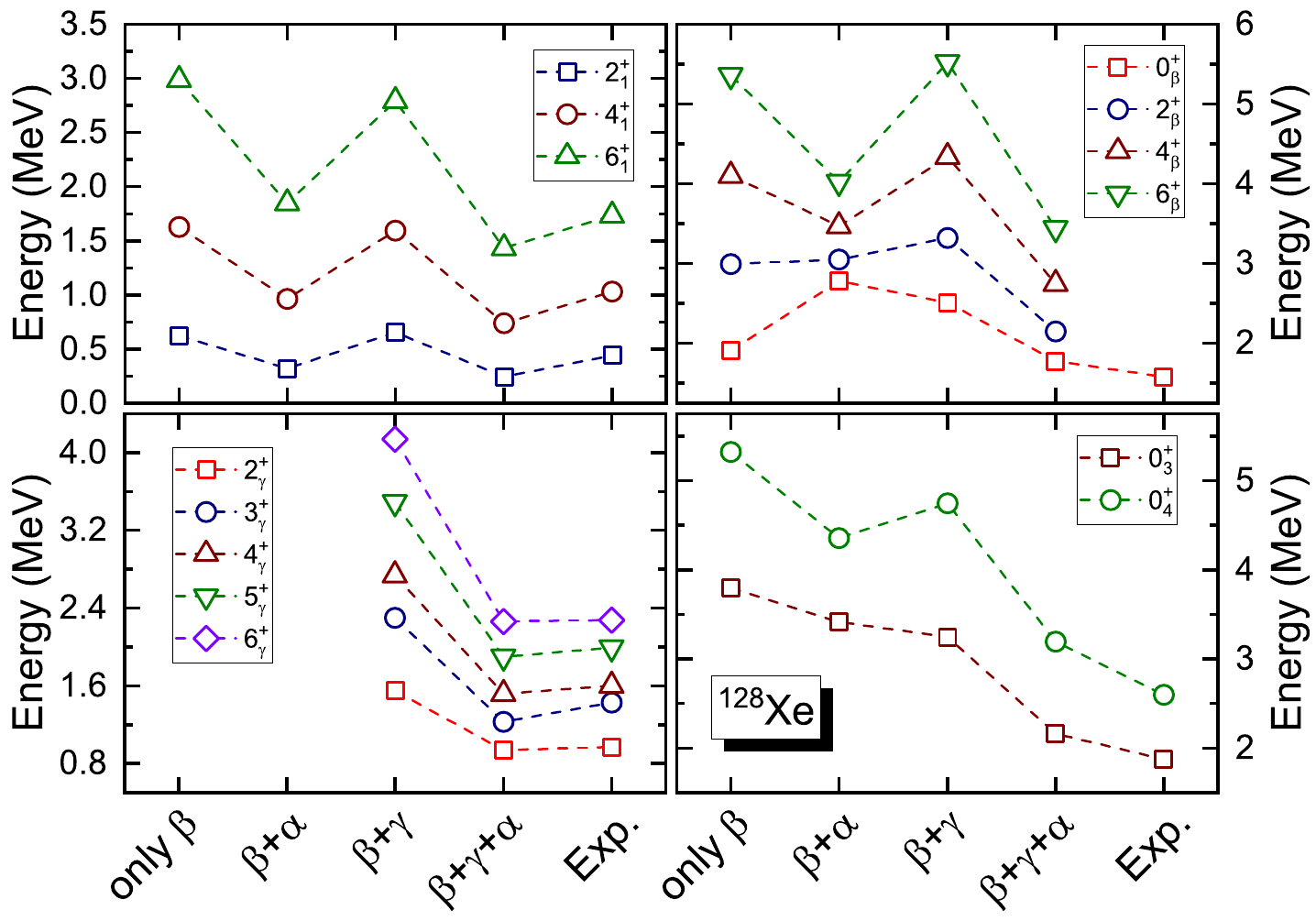}}
  \caption{\label{lowenergy}(Color online) Excitation energies of the g.s. band, the band based on $0^+_\beta$ ($0^+_2$), the $\gamma$-band, and the $0^+_{3,4}$ states in $^{128}$Xe. Results obtained with the collective Hamiltonian including one-dimensional axial-quadrupole ($\beta$), axial-quadrupole plus pairing ($\beta+\alpha$), triaxial-quadrupole ($\beta+\gamma$), and triaxial-quadrupole plus pairing ($\beta+\gamma+\alpha$) degrees of freedom are compared with the corresponding experimental values~\cite{NNDC}. }
  \end{figure}

  \begin{figure}[ht]
    \centering{\includegraphics[width=0.45\textwidth]{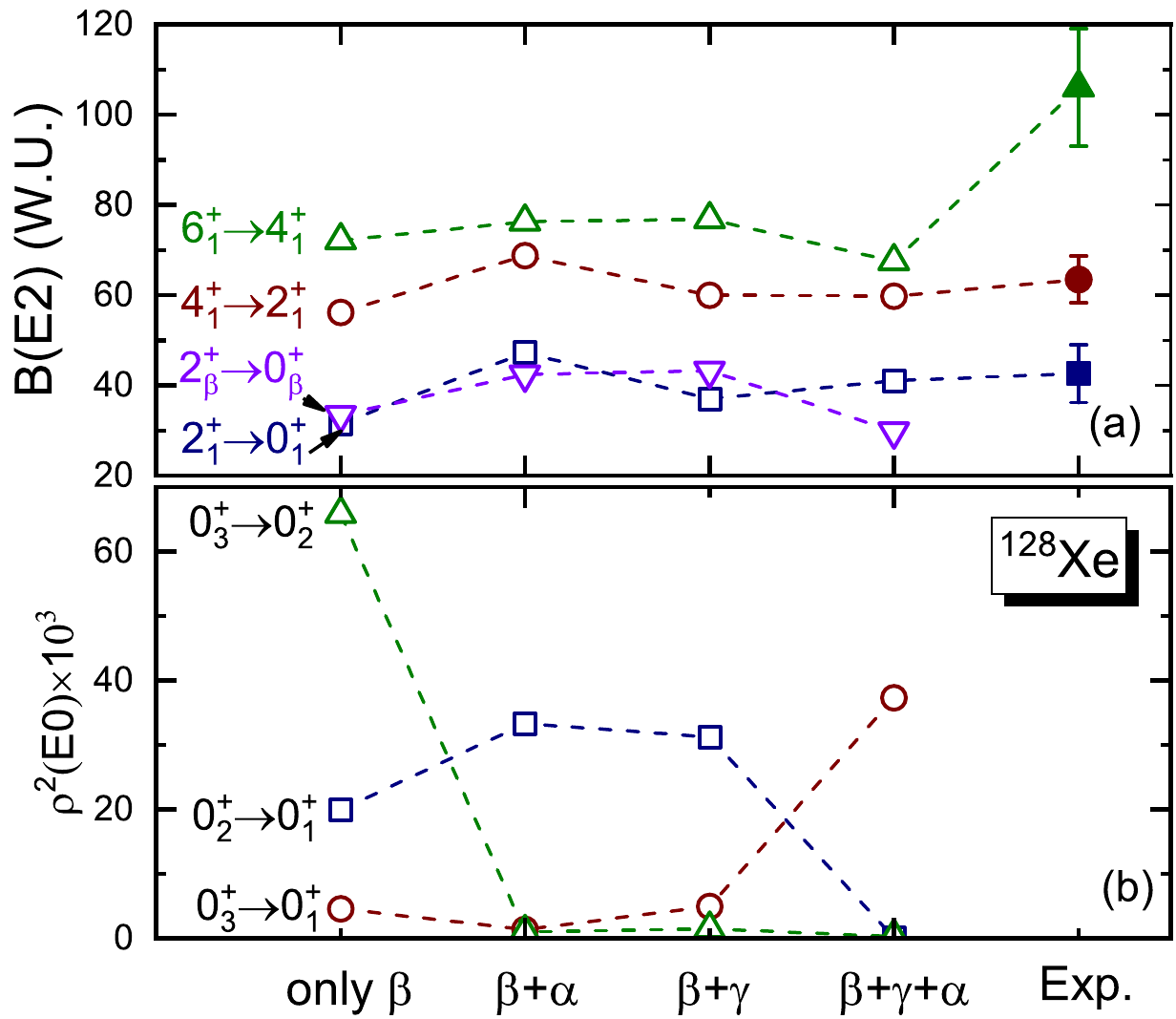}}
    \caption{\label{BE2-E0}(Color online) Same as in the caption to Fig. \ref{lowenergy}, but for the intra-band $B(E2)$ values in the ground-state band, and the $0^+_\beta$ band of $^{128}$Xe (a). The $E0$ transition strengths between $0^+$ states are shown in the lower panel (b).}
    \end{figure}

The diagonalization of the resulting Hamiltonian yields the energy spectra and collective wave functions for each value of the total angular momentum. Fig.~\ref{lowenergy} displays the excitation energies of the ground-state (g.s.) band, the excited band based on $0^+_\beta$, the $\gamma$-band, and the $0^+_{3,4}$ states in $^{128}$Xe. Results obtained with the collective Hamiltonian that includes the one-dimensional axial-quadrupole ($\beta$), axial-quadrupole plus pairing ($\beta+\alpha$), triaxial-quadrupole ($\beta+\gamma$)~\cite{NiksicPRC2009}, and triaxial-quadrupole plus pairing ($\beta+\gamma+\alpha$) degrees of freedom, are compared with experimental values~\cite{NNDC}. Obviously, the coupling between shape and pairing dynamical degrees of freedom has a pronounced effect on the calculated spectra, especially for the triaxial calculation. The inclusion of triaxial deformation and pairing vibrational degrees of freedom generally lowers the low-lying energy spectra, bringing the excitation energies in a quantitatively much better agreement with experiment, especially for the $\gamma$-band and the excited $0^+$ states. As shown in Fig. \ref{BE2-E0}, the inclusion of the dynamical pairing degree of freedom has limited effect on the calculated intra-band $B(E2)$ values, and all theoretical results appear to be consistent with the data. For the $E0$ transitions, one finds that the $\rho^2(E0; 0^+_2\rightarrow 0^+_1)$ and $\rho^2(E0; 0^+_3\rightarrow 0^+_1)$ invert values, and $\rho^2(E0; 0^+_3\rightarrow 0^+_2)$ almost vanishes by adding dynamical pairing to the triaxial collective Hamiltonian. This indicates a different structure of $0^+_2$ and $0^+_3$ states, that interchange when including dynamical pairing, as shown by the probability density distributions calculated from the collective wave functions (c.f. Eq. \ref{eq:rho-nI}).

\begin{figure}[ht]
  \centering{\includegraphics[width=0.5\textwidth]{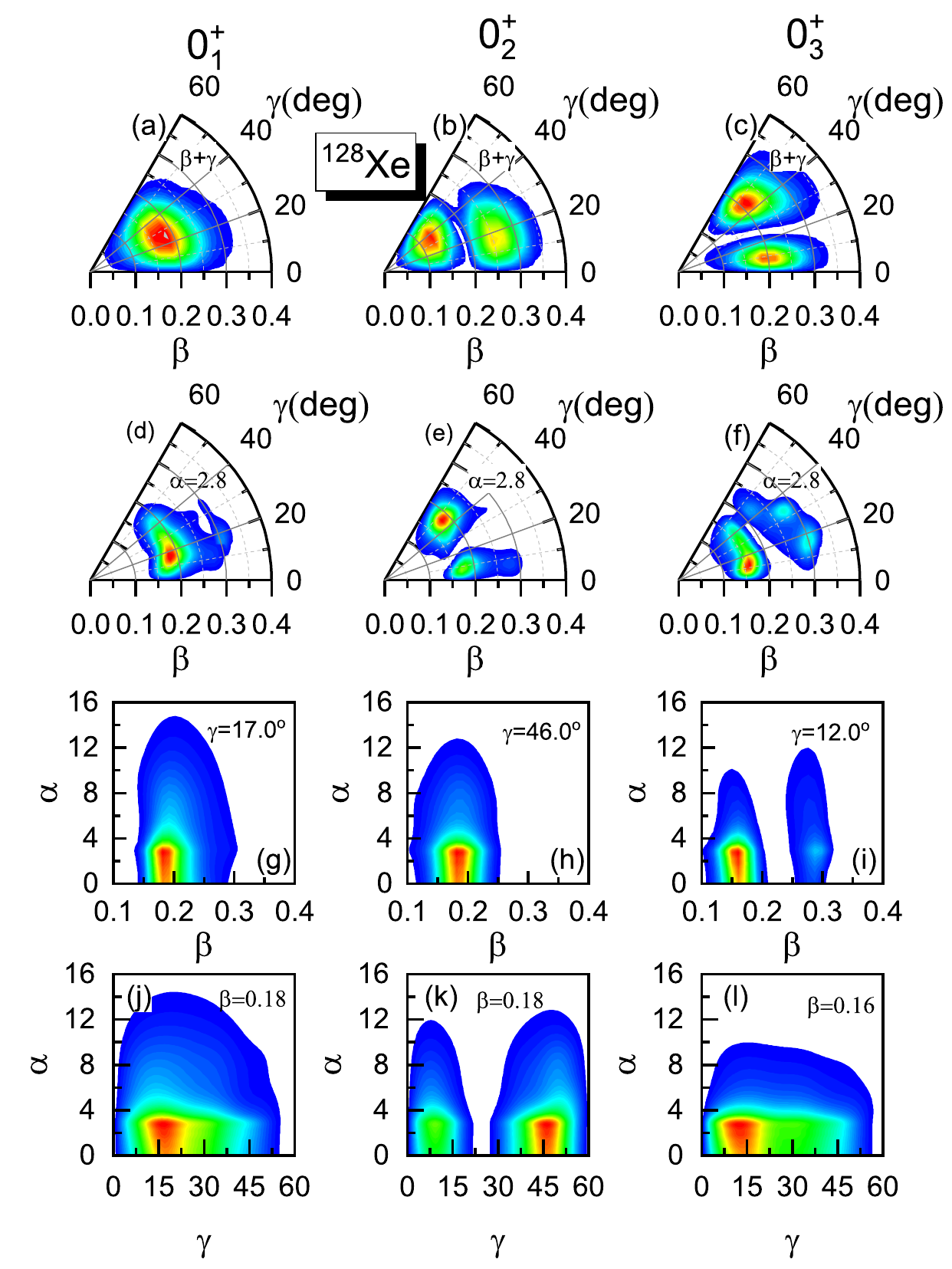}}
  \caption{\label{wav}(Color online) Probability density distributions in the $(\beta, \gamma)$ plane for the first three $0^+$ states of $^{128}$Xe calculated with the triaxial-quadrupole collective Hamiltonian (a-c). Projections of the probability density distributions in the $(\beta, \gamma)$, $(\beta,\alpha)$, and $(\gamma, \alpha)$ planes, with the third axis fixed at the peak position in $(\beta,\gamma,\alpha)$ calculated with the TPCH (d-l).}
\end{figure}

Fig.~\ref{wav} displays the projections of the probability density distributions in the $(\beta, \gamma)$, $(\beta, \alpha)$, and $(\gamma, \alpha)$ planes for the first three $0^+$ states of $^{128}$Xe, calculated with the TPCH. The corresponding third axis is fixed at the peak position in $(\beta,\gamma,\alpha)$, and these values are also shown in the two-dimensional plots. For comparison, the probability density distributions calculated with the triaxial-quadrupole collective Hamiltonian are plotted in the upper row. The peak of the probability density distribution for $0^+_1$ is found at $(\beta,\ \gamma,\ \alpha)=(0.18,\ 17^\circ,\ 2.8)$, whereas the peak of $0^+_2$ is located at $(0.18,\ 46^\circ,\ 2.8)$. The large difference in  the triaxial degree of freedom leads to a small overlap between $0^+_1$ and $0^+_2$, and consequently a small value of $\rho^2(E0; 0^+_2\rightarrow 0^+_1)$. In contrast, the peak of the probability density distribution of $0^+_3$ is close to that of the ground state, and results in a considerable $E0$ transition. In the TPCH calculation, nodes are found in the $\gamma$ direction for $0^+_2$, and $\beta$ direction for $0^+_3$, opposite to those obtained with the triaxial collective Hamiltonian. This is the reason for the interchange of $\rho^2(E0; 0^+_2\rightarrow 0^+_1)$ and $\rho^2(E0; 0^+_3\rightarrow 0^+_1)$ that occurs by including dynamical pairing in the triaxial calculation (c.f. Fig. \ref{BE2-E0}).

\begin{figure*}[ht]
\centering{\includegraphics[width=0.90\textwidth]{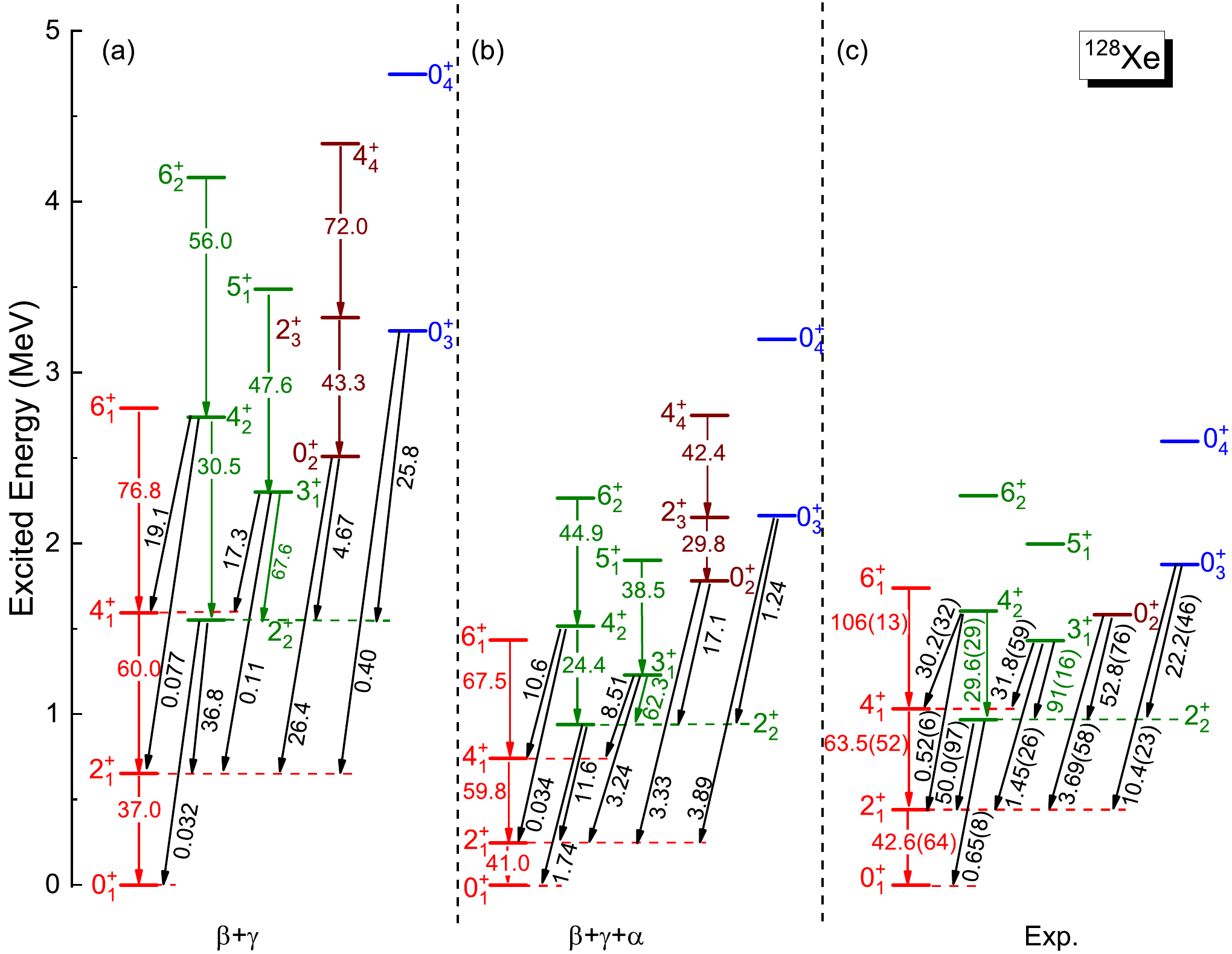}}
\caption{\label{Spectra-Xe128}(Color online) Low-energy excitation spectra and electric quadrupole transitions ($B(E2)$ values in Weisskopf units) for $^{128}$Xe, calculated with the triaxial-quadrupole collective Hamiltonian($\beta+\gamma$) (a), and the TPCH ($\beta+\gamma+\alpha$) (b), based on the PC-PK1 energy density functional~\cite{ZhaoPRC2010}. The theoretical values are shown in comparison with the available data (c) from Ref.~\cite{NNDC}.}
\end{figure*}

We further demonstrate that the model is also capable of describing detailed structure properties of $\gamma$-soft nuclei. Fig.~\ref{Spectra-Xe128} displays the excitation spectra for $^{128}$Xe calculated with the triaxial-quadrupole collective Hamiltonian ($\beta+\gamma$) and TPCH ($\beta+\gamma+\alpha$), in comparison with available data~\cite{NNDC}.
In general, the energy spectrum is compressed by including the dynamical pairing degree of freedom, and in very good agreement with experiment, especially for the $\gamma$-band and the excited $0^+$ states. This is because the inclusion of dynamical pairing increases the moments of inertia and collective masses \cite{XiangPRC2020}. One also notices that the inter-band $B(E2)$ values for transitions from $0^+_2$ to the first two $2^+$ states, and from $0^+_3$ to $2^+_1$, are significantly improved because of the interchange of the two excited $0^+$ states (c.f. Fig. \ref{wav}) when dynamical pairing is included. The inter-band transitions between the $\gamma$-band and g.s. band are generally reduced, which could probably be related to the compression of the g.s. band caused by the enhanced moments of inertia. This may be improved by including pairing rotation in the present TPCH \cite{PomorskiIJMPE2007}.

\subsection{The effect of ZPE on low-lying spectra}

\begin{figure}[ht]
  \centering{\includegraphics[width=0.45\textwidth]{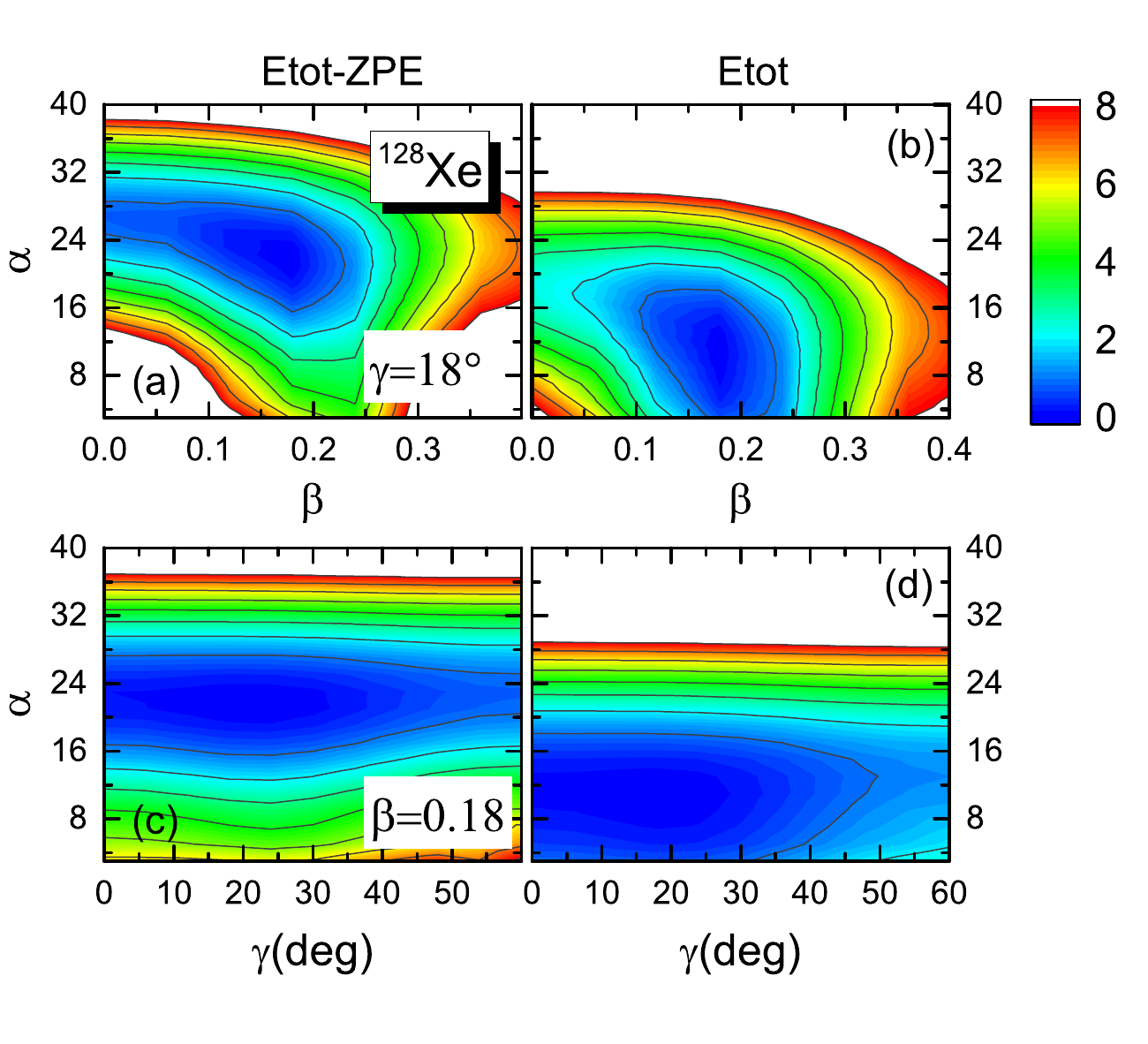}}
  \caption{\label{Vc-Etot}(Color online) Projections of the PESs of $^{128}$Xe with (a,c) and without (b,d) ZPE, computed by the constrained CDFT based on the PC-PK1 functional. The two-dimensional projections of the PESs are  shown in the $(\beta, \alpha)$ and $(\gamma, \alpha)$ planes with the fixed third components corresponding to the global minimum. }
  \end{figure}

The zero-point energy (ZPE) correction often plays a crucial role in low-energy excitation spectra by modifying the topology of mean-field potential energy surfaces. Here we illustrate the effect of the ZPE on PESs and low-lying spectra. Fig.~\ref{Vc-Etot} displays the projections of the PESs of $^{128}$Xe with and without ZPE, computed by the constrained CDFT based on the PC-PK1 functional. The two-dimensional projections of the PESs are  shown in the $(\beta, \alpha)$ and $(\gamma, \alpha)$ planes with the fixed third components corresponding to the global minimum. Obviously, the topology of the collective potential is significantly modified by the inclusion of ZPE, and the minimum is shifted toward larger  values of $\alpha$. Consequently, the moments of inertia are reduced \cite{XiangPRC2020} and the excitation spectra extend. This is also shown in Tab.~\ref{Tab-Q2}, where the excitation energies of low-lying states of $^{128}$Xe, calculated with the TPCH without (W/O) and with (W/ZPE) ZPE, are tabulated in comparison with available data~\cite{NNDC}. One finds that the excitation energies increase by $~18\%-50\%$, and are in better agreement with the data when ZPE corrections are included.

\begin{table}[!htbp]
\tabcolsep=5pt   
\begin{center}
\caption{\label{Tab-Q2}  Excitation energies of low-lying states of $^{128}$Xe calculated using the TPCH without (W/O) and with ZPE (W/ZPE), in comparison with available data~\cite{NNDC}.}
\begin{tabular}{cccc}
\hline\hline
\ & W/O(MeV)  &  W/ZPE(MeV) & Exp.(MeV)\\
\hline
$2^+_1$     &  0.202   &  0.247      & 0.443 \\
$4^+_1$     &  0.608   &  0.740      & 1.033 \\
$6^+_1$     &  1.191   &  1.434      & 1.737 \\
$0^+_\beta$ &  1.240   &  1.781      & 1.583 \\
$2^+_\beta$ &  1.572   &  2.152      &   -   \\
$4^+_\beta$ &  1.802   &  2.749      &   -   \\
$2^+_\gamma$&  0.770   &  0.940      & 0.970\\
$3^+_\gamma$&  1.051   &  1.229      & 1.430\\
$4^+_\gamma$&  1.275   &  1.517      & 1.604\\
$5^+_\gamma$&  1.658   &  1.900      & 1.997\\
$6^+_\gamma$&  1.905   &  2.265      & 2.281\\
$0^+_3$     &  1.735   &  2.162      & 1.877 \\
$0^+_4$     &  2.651   &  3.196      & 2.598 \\

\hline\hline
\end{tabular}
\end{center}
\end{table}


\section{SUMMARY}\label{Sec:IIII}

The triaxial quadrupole collective Hamiltonian, based on relativistic energy density functionals, has been extended to include a pairing collective coordinate, and is here referred as TPCH . In addition to triaxial shape vibrations and rotations, the TPCH describes pairing vibrations and the coupling between triaxial shape and pairing degrees of freedom. The parameters of the collective Hamiltonian are fully determined by constrained CDFT calculations in the space of intrinsic triaxial shape and pairing deformations $(\beta, \gamma, \alpha)$. The effect of coupling between triaxial shape and pairing degrees of freedom has been analyzed in a study of low-lying excitation spectra and transition rates of $^{128}$Xe. When compared to results obtained with the standard triaxial collective Hamiltonian, the inclusion of the dynamical pairing degree of freedom compresses the low-lying spectra, in better agreement with data. Furthermore, the structure of the $0^+_2$
and $0^+_3$ states is exchanged with dynamical pairing, and this modifies significantly the $E0$ transition strengths between $0^+$ states. Remarkably, the inter-band $B(E2)$ transition probabilities from excited $0^+$ states to $2^+$ states have been improved by including dynamical pairing. Finally, we have shown that the inclusion of ZPE can alter the topology of the PES and shift the global minimum to larger values of $\alpha$, thus increasing the excitation energies by $~18\%-50\%$.

\begin{acknowledgments}
This work has been supported in part by the National Natural Science Foundation of China under Grants No. 12005109,  No. 12005082, No. 12375126, the PHD Foundation of Chongqing Normal University (No. 23XLB010), the Science and Technology Research Program of Chongqing Municipal Education Commission (No. KJQN202300509), the Fundamental Research Funds for the Central Universities, the QuantiXLie Centre of Excellence, a project co-financed by the Croatian Government and European Union through the European Regional Development Fund - the Competitiveness and Cohesion Operational Programme (KK.01.1.1.01.0004).
\end{acknowledgments}

\appendix
\section*{APPENDIX: Derivation of collective masses and zero-point energies \label{App:BZPE}}

Following Ref.~\cite{BaranPRC2011}, we present the derivation of collective mass and zero-point energy correction, including the $\beta$, $\gamma$, and $\alpha$ degrees of freedom. In the adiabatic time-dependent Hartree-Fock-Bogoliubov (ATDHFB) method, a generalized density matrix is expanded around the quasi-stationary HFB solution $\mathcal{R}_0$ up to quadratic terms in the collective momentum:
\begin{align}
\mathcal{R}=\mathcal{R}_0+\mathcal{R}_1+\mathcal{R}_2
\end{align}
where $\mathcal{R}_1$ is time-odd, and $\mathcal{R}_0$ and $\mathcal{R}_2$ are time-even densities.
The corresponding expansion for the HFB Hamiltonian matrix takes the form
\begin{align}
\mathcal{W}=\mathcal{W}_0+\mathcal{W}_1+\mathcal{W}_2.
\end{align}
In ATDHFB, the general form of collective mass reads
\begin{align}
B&=\frac{i}{2\dot{q}^2}\text{Tr}\lrb{\dot{\mathcal{R}}_0\lrs{\mathcal{R}_0,\mathcal{R}_1}}\nonumber\\
&=\frac{i}{2\dot{q}}\text{Tr}\lrb{\frac{\partial\mathcal{R}_0}{\partial{q}}\lrs{\mathcal{R}_0,\mathcal{R}_1}}
\end{align}
where $q$ denotes the collective coordinate. The trace in the expression above can easily be evaluated in the quasiparticle basis. To this end, one can utilize the ATDHFB equation
\begin{align}\label{eq-idotR_0}
i\dot{\mathcal{R}}_0=\lrs{\mathcal{W}_0,\mathcal{R}_1}+\lrs{\mathcal{W}_1,\mathcal{R}_0}.
\end{align}
In the quasiparticle basis, the matrices $\mathcal{R}_0$, $\mathcal{W}_0$, $\mathcal{W}_1$, $\mathcal{R}_1$, and
$\dot{\mathcal R}_0$ are expressed in terms of the matrices $\mathcal{G}$, $\mathcal{E}_0$, $\mathcal{E}_1$, $\mathcal{Z}$, and $\mathcal{F}$, respectively:
\begin{align}\label{eq-R_0}
\mathcal{R}_0=\mathcal{A}^\dagger\mathcal{G}\mathcal{A}\\
\mathcal{W}_0=\mathcal{A}^\dagger\mathcal{E}_0\mathcal{A}\\
\mathcal{W}_1=\mathcal{A}^\dagger\mathcal{E}_1\mathcal{A}\\
\mathcal{R}_1=\mathcal{A}^\dagger\mathcal{Z}\mathcal{A}\\
\dot{\mathcal{R}}_0=\mathcal{A}^\dagger\mathcal{F}\mathcal{A}\label{eq-dotR0}
\end{align}
where
\begin{align}\label{eq-Bogtrans-A}
\mathcal{A}
=\lrb{\begin{array}{cc}
U & V^*\\
V & U^*
\end{array}}
\end{align}
is the matrix of Bogoliubov transition, and
\begin{align}
\mathcal{G}=\lrb{\begin{array}{cc}
0 & 0\\
0 & 1
\end{array}},\ \ \
\mathcal{E}_0=\lrb{\begin{array}{cc}
E & 0\\
0 & -E
\end{array}}.
\end{align}
Inserting Eqs. (\ref{eq-R_0}-\ref{eq-Bogtrans-A}) into (\ref{eq-idotR_0}), one obtains
\begin{align}
i\mathcal{F}=\lrs{\mathcal{E}_0,\mathcal{Z}}+\lrs{\mathcal{E}_1,\mathcal{G}}
\end{align}
This $2\times2$ matrix equation is equivalent to the following equation
\begin{align}\label{eq-iF1}
iF = EZ + ZE +E_1,
\end{align}
with the relations
\begin{align}
\mathcal{F}=\lrb{\begin{array}{cc}
0 & F^*\\
F & 0
\end{array}}\\
\mathcal{Z}=\lrb{\begin{array}{cc}
0  & Z^*\\
Z & 0
\end{array}}\\
\lrs{\mathcal{E}_1,\mathcal{G}}=\lrb{\begin{array}{cc}
0 & E_1^*\\
E_1 & 0
\end{array}}.
\end{align}
In the case of multiple collective coordinates $\{q_i\}$, the ATDHFB equation (\ref{eq-idotR_0}) needs to be solved for each coordinate
\begin{align}\label{eq-idotR_0i}
i\dot{q}_i\frac{\partial{\mathcal{R}}_0}{\partial{q_i}}=\lrs{\mathcal{W}_0,\mathcal{R}^i_1}+\lrs{\mathcal{W}_1^i,\mathcal{R}_0},
\end{align}
and the collective mass tensor becomes
\begin{align}
B_{ij}=\frac{i}{2\dot{q}_j}
\text{Tr}\lrb{\frac{\partial\mathcal{R}_0}{\partial{q_i}}\lrs{\mathcal{R}_0,\mathcal{R}^j_1}}.
\end{align}
Then, in terms of the corresponding matrices $F^i$ and $Z^j$, the collective mass tensor is given by the expression
\begin{align}\label{eq-CBFZ}
B_{ij}&=\frac{i}{2\dot{q}_i\dot{q}_j}\text{Tr}\lrb{F^{i*}Z^j-F^iZ^{j*}}\nonumber\\
&=\frac{i}{2\dot{q}_i\dot{q}_j}\text{Tr}\lrb{\sum\limits_{\nu}\lrb{F^{i*}_{\mu\nu}Z^j_{\nu\mu}
-F^i_{\mu\nu}Z^{j*}_{\nu\mu}}}.
\end{align}
Since $Z$ is an antisymmetric matrix, one obtains
\begin{align}\label{eq-CBFZ2}
B_{ij}&=-\frac{i}{2\dot{q}_i\dot{q}_j}\text{Tr}\lrb{\sum\limits_{\nu}\lrb{F^{i*}_{\mu\nu}Z^j_{\mu\nu}
-F^i_{\mu\nu}Z^{j*}_{\mu\nu}}}\nonumber\\
&=-\frac{i}{2\dot{q}_i\dot{q}_j}\sum\limits_\mu\sum\limits_{\nu}\lrb{F^{i*}_{\mu\nu}Z^j_{\mu\nu}
-F^i_{\mu\nu}Z^{j*}_{\mu\nu}}.
\end{align}
In most ATDHFB applications, the time-odd interaction matrix $E_1$
appearing in Eq. (\ref{eq-iF1}) has been neglected. In the following, this approximation will be referred to as the cranking approximation. Without the $E_1$ term, the matrix $Z$ can be obtained in the quasiparticle basis from the equation
\begin{align}
iF^i_{\mu\nu}=\lrb{E_\mu+E_\nu}Z^i_{\mu\nu},
\end{align}
and
\begin{align}
B_{ij}=\frac{1}{2\dot{q}_i\dot{q}_j}\sum\limits_{\mu\nu}\frac{F^{i*}_{\mu\nu}F^j_{\mu\nu}+F^j_{\mu\nu}F^{j*}_{\mu\nu}}{E_\mu+E_\nu}.
\end{align}
Eq.~(\ref{eq-dotR0}) can explicitly be written in terms of HFB eigenvectors:
\begin{align}
\dot{\mathcal{R}_0}
&=\dot{q}\frac{\partial}{\partial{q}}\lrb{
\begin{array}{cc}
\rho_0  & \kappa_0\\
-\kappa^*_0 & 1-\rho^*_0
\end{array}}\nonumber\\
&=\lrb{\begin{array}{cc}
UFV^T-V^*F^*U^\dagger & UFU^T-V^*F^*V^\dagger\\
VFV^T-U^*F^*U^\dagger & VFU^T-U^*F^*V^\dagger
\end{array}}.
\end{align}
Evaluating the matrix elements in the canonical basis, we obtain
\begin{align}
F^i_{\mu\bar{\nu}}=\frac{s_{\bar{\nu}}}{u_\mu{v}_\nu+v_\mu{u}_\nu}\dot{q}_i\lrb{\frac{\partial\rho_0}{\partial{q}_i}}_{\mu\nu} .
\end{align}
By differentiating the HFB equation $\lrs{\mathcal{W}_0,\mathcal{R}_0}=0$ with respect to $q_i$ , the derivative of the density matrix can be expressed in terms of the derivatives of the particle-hole and pairing mean fields
\begin{align}
\lrs{\mathcal{A}^\dagger\dot{q}_i\frac{\partial{\mathcal{W}_0}}{\partial{q}_i}\mathcal{A},\mathcal{G}}
+\lrs{\mathcal{E},\mathcal{F}}=0.
\end{align}
The following expression is obtained obtained
\begin{align}
F_{\mu\nu}=\frac{\dot{q}_i\lrb{U^\dagger\frac{\partial{h}}{\partial{q_i}}V^*-V^\dagger\frac{\partial{\Delta^*}}{\partial{q_i}}V^*
+U^\dagger\frac{\partial{\Delta}}{\partial{q_i}}U^*-V^\dagger\frac{\partial{h^*}}{\partial{q_i}}U^*}_{\mu\nu}}{E_\mu+E_\nu},
\end{align}
and then, in the canonical basis $B_{\mu\nu}$ reads
\begin{align}
B_{\mu\nu}=\frac{\hbar^2}{2}\sum\limits_{\tau\iota}\frac{f^*_{\mu,\tau\iota}f_{\nu,\tau\iota}
+f_{\mu,\tau\iota}f^*_{\nu,\tau\iota}}{E_\tau+E_\iota},
\end{align}
with
\begin{align}\label{eq-fhdel}
f_{\mu,\tau\iota}&=\frac{1}{E_\tau+E_\iota}\left[-\lrb{u_{\tau}v_{\iota}+v_{\tau}u_\iota}\lrb{\partial_\mu h_0}_{{\bar\tau}\iota}\right.\nonumber\\
&\left.+u_{\tau}u_{\iota}\lrb{\partial_\mu\Delta}_{\tau\iota}
+v_{\tau}v_{\iota}\lrb{\partial_\mu\Delta^\dagger}_{\bar\tau\bar\iota}\right],
\end{align}
where
\begin{align}
\lrb{h_0}_{{\bar\tau}\iota}=\langle\Phi\mid\left\{\lrs{a_{\bar{\tau}},H},a^\dagger_\iota\right\}\mid\Phi\rangle\\
\Delta_{\tau\iota}=\langle\Phi\mid\left\{\lrs{a_\tau,H},a_{\iota}\right\}\mid\Phi\rangle\\
\Delta_{\bar\tau\bar\iota}^\dagger=\langle\Phi\mid\left\{a_{\bar\iota}^\dagger,\lrs{H,a_{\bar\tau}^\dagger}\right\}\mid\Phi\rangle .
\end{align}

In the present study, the set of collective coordinates $q_i$ is $\{\beta,\ \gamma,\ \alpha\}$, and the intrinsic collective states are obtained in a self-consistent mean-field calculation, with additional constraints on the expectation value of the Hamiltonian:
\begin{align}
&\langle H\rangle+\sum\limits_{\mu=0,2}\lambda_{2\mu}\left(\langle \hat{Q}_{2\mu}\rangle - q_{2\mu}\right)\nonumber\\
&-\lambda\langle\hat{N}-N\rangle
-\lambda_\alpha\langle\hat{P}-\alpha\rangle,
\label{constr2}
\end{align}
where the pairing operator $\hat{P}$ is defined
\begin{align}
\hat{P}=\frac{1}{2}\sum\limits_{\mu>0}\lrb{a_\mu^\dagger a_{\bar\mu}^\dagger+a_{\bar\mu}a_\mu},\ \ \
\text{and}\ \ \
\hat{P}^\dagger=\hat{P}.
\end{align}
The expectation value of $\hat{P}$ in the intrinsic collective state $|\Phi\rangle$ is $\alpha$ by definition:
\begin{align}
\alpha=\langle\Phi\mid\hat{P}\mid\Phi\rangle=\sum\limits_{\mu>0}u_\mu v_\mu
\end{align}
and, therefore, $\partial_{\lambda_\alpha}\Delta_{\tau\iota}$ can be rewritten as
\begin{align}
\partial_{\lambda_\alpha}\Delta_{\tau\iota}
&=-\frac{1}{2}\sum_{i>0}\lrb{\delta_{\tau i}\delta_{\bar{i}\iota}-\delta_{\tau \bar{i}}\delta_{i\iota}}\label{eq-pDlamalp}\\
\partial_{\lambda_\alpha}\Delta_{\bar\tau\bar\iota}^\dagger
&=-\frac{1}{2}\sum_{i>0}\lrb{\delta_{\bar\tau i}\delta_{\bar{i}\bar\iota}-\delta_{\bar\tau \bar{i}}\delta_{i\bar\iota}}\label{eq-pDlamalpdia}.
\end{align}
Inserting Eqs. (\ref{eq-pDlamalp}, \ref{eq-pDlamalpdia}) into Eq. \ref{eq-fhdel}, one obtains
\begin{align}
f_{\lambda_\alpha,\tau\iota}&=-\frac{1}{2}\sum_{i>0}\frac{{\lrb{u_{\tau}u_{\iota}-v_{\tau}v_{\iota}}}}{E_\tau+E_\iota}
\lrb{\delta_{\tau i}\delta_{{i}\bar\iota}-\delta_{\bar\tau{i}}\delta_{\bar{i}\bar{\iota}}},
\end{align}
and also
\begin{align}
B_{\lambda_\alpha\lambda_\alpha}
&=\frac{1}{4}\sum\limits_{\tau\iota}
\frac{\lrb{u_{\tau}u_{\iota}-v_{\tau}v_{\iota}}^2}{\lrb{E_\tau+E_\iota}^3}
\delta_{\tau\bar{\iota}}\;.
\end{align}
For $B_{\lambda_\alpha\lambda_{2\mu}}$, we also need
\begin{align}
f_{\lambda_{2\mu},\tau\iota}
&=-\frac{1}{E_\tau+E_\iota}\lrs{\lrb{u_{\tau}v_{\iota}
+v_{\tau}u_\iota}\lrb{\partial_{\lambda_{2\mu}} h_0}_{\bar{\tau}\iota}} \\
&=\frac{1}{E_\tau+E_\iota}\lrs{u_{\tau}v_{{\iota}}+v_{\tau}u_\iota }Q_{2\mu,\bar{\tau}\iota}.
\end{align}
Then, the following expressions are obtained
\begin{align}
  B_{\lambda_\alpha\lambda_{2\mu}}
  &=-\hbar^2\sum_{i>0}\frac{\lrb{u_{i}^2-v_{i}^2}\lrb{u_{i}v_{i}+v_{i}u_i}Q_{2\mu,ii}}{\lrb{E_i+E_i}^3}
\end{align}
and
\begin{align}
B_{\lambda_{2\mu}\lambda_{2\nu}}
&=\hbar^2\sum\limits_{ij}\frac{Q_{2\mu,i\bar{j}}Q_{2\nu,\bar{j}i}\lrb{u_{i}v_{j}+v_{i}u_j }^2}{\lrb{E_i+E_j}^3}.
\end{align}
Furthermore,
\begin{align}
B_{ab}=\sum\limits_{mn}\frac{\partial\lambda_n}{\partial{a}}\frac{\partial\lambda_n}{\partial{b}}B_{\lambda_m\lambda_n}\;,
\end{align}
where $a,b,m,n$ denote $q_{20},q_{22}$ and $\alpha$. The partial derivatives can be calculated perturbatively
\begin{align}
\delta\mid\Phi\rangle
=-\delta\lambda_\alpha\sum\limits_{k<k'}\frac{\langle\Phi\mid\beta_{k'}\beta_k\hat{P}\mid\Phi\rangle}{E_{k'}+E_k}
\beta_{k}^\dagger\beta_{k'}^\dagger\mid\Phi\rangle\  ,
\end{align}
and
\begin{align}
\delta\alpha
&=\delta\langle\Phi\mid\hat{P}\mid\Phi\rangle\nonumber\\
&=- 2 \delta\lambda_\alpha \sum\limits_{k<k'}\frac{\langle\Phi\mid\beta_{k'}\beta_k\hat{P}\mid\Phi\rangle}{E_{k'}+E_k}
\langle\Phi\mid\hat{P}^\dagger\beta_{k}^\dagger\beta_{k'}^\dagger\mid\Phi\rangle .
\end{align}
The partial derivative then reads
\begin{align}
\frac{\delta\alpha}{\delta\lambda_\alpha}
&=- 2 \sum\limits_{k<k'}\frac{\mid\langle\Phi\mid\beta_{k'}\beta_k\hat{P}\mid\Phi\rangle\mid^2}{E_{k'}+E_k}, \nonumber
\end{align}
where
\begin{align}
\langle\Phi\mid\beta_{k'}\beta_k\hat{P}\mid\Phi\rangle
&=-\frac{1}{2}\lrb{u^2_{k'}-v^2_{k'}}\delta_{\bar kk'}(k'>0)
\end{align}
Finally, the expression for the partial derivative
\begin{align}\label{eq-partalp-lamba}
\frac{\delta\alpha}{\delta\lambda_\alpha}
=-\sum_{k'>0}\frac{\lrb{u^2_{k'}-v^2_{k'}}^2}{4E_{k'}} .
\end{align}
Similarly, the other derivatives read
\begin{align}
\frac{\delta{q_{2\mu}}}{\delta\lambda_{2\nu}}
&=-\sum\limits_{k,k'}\frac{Q_{2\mu,k\bar{k'}}Q_{2\nu,\bar{k'}k}\lrb{u_kv_{k'}+u_{k'}v_k}^2}{E_{k'}+E_k}
\end{align}

\begin{align}
\frac{\delta\alpha}{\delta\lambda_{q_{2\mu}}}
&=\sum\limits_{k'>0}\frac{Q_{2\mu,\bar{k'}\bar{k'}}\lrb{u_{k'}v_{k'}+u_{k'}v_{k'}}}{E_{k'}+E_{k'}}\lrb{u^2_{k'}-v^2_{k'}}
\end{align}
\begin{align}
\frac{\delta{q_{2\mu}}}{\delta\lambda_\alpha}
&=\sum\limits_{k'>0}\frac{Q_{2\mu,\bar{k'}\bar{k'}}\lrb{u_{k'}v_{k'}+u_{k'}v_{k'}}}{E_{k'}+E_{k'}}\lrb{u^2_{k'}-v^2_{k'}}
\end{align}

The mass parameters are thus determined by the following expressions
\begin{align}
B_{ab}=\hbar^2\lrs{\mathcal{M}^{-1}_{(1)}\mathcal{M}_{(3)}\mathcal{M}^{-1}_{(1)}}_{ab} ,
\end{align}
with
\begin{align}\label{eq-metrixM}
\mathcal{M}_{(n)ab}=\sum\limits_{ij}\frac{\mathcal{O}^\mu_{ij}\mathcal{O}^\nu_{ji}}{\lrb{E_i+E_j}^n} ,
\end{align}
and
\begin{align}
\mathcal{O}^{q_{2\mu}}_{ij}&=\left\langle i\right|\hat{Q}_{2\mu}\left| j\right\rangle
\left(u_i v_j+ v_i u_j \right)\\
\mathcal{O}^\alpha_{ij}&=-\frac{1}{2}\lrb{u^2_{i}-v^2_{i}}\delta_{ij} .
\end{align}

The Zero-Point Energy (ZPE) is calculated in the cranking approximation, that is, on the same level of
approximation as the mass parameters and moments of inertia.
The vibrational ZPE is given by the expression~\cite{GirodNPA1979}
\begin{align}
\Delta{V}_{\rm vib}=\frac{1}{2}\text{\textbf{Tr}}\lrs{B^{-1}G} .
\end{align}
Similar to the derivation of the collective mass, an expression for the vibrational ZPE can be obtained
\begin{align}
\Delta{V}_{\text vib}=\frac{1}{4}\text{Tr}\lrb{\mathcal{M}^{-1}_{(3)}\mathcal{M}_{2}}_{ab} ,
\end{align}
where the matrix $\mathcal{M}$ is defined in Eq.~(\ref{eq-metrixM}).

\bibliography{reference}

\end{document}